\documentclass[twocolumn,superscriptaddress]{revtex4-1}
\usepackage{eurosym}
\usepackage{amsmath}
\usepackage{graphicx}
\usepackage{bm}
\usepackage{color}
\usepackage{xcolor}
\usepackage{ulem}

\def  \bsig   {\mbox{\boldmath$\sigma $}}

\begin{document}

\title{Magnetic scattering with spin-momentum locking: Single scatterers and diffraction grating}

\author{S. Wolski}
\affiliation{Department of Physics and Medical Engineering, Rzesz\'ow University of
Technology, al.~Powsta\'nc\'ow Warszawy 6, 35-959 Rzesz\'ow, Poland}

\author{V. K. Dugaev}
\affiliation{Department of Physics and Medical Engineering, Rzesz\'ow University of
Technology, al.~Powsta\'nc\'ow Warszawy 6, 35-959 Rzesz\'ow, Poland}

\author{E. Ya. Sherman}
\affiliation{Department of Physical Chemistry, University of the Basque Country, 48940,
Leioa, Spain}
\affiliation{Ikerbasque, Basque Foundation for Science, Bilbao, Spain}
\affiliation{EHU Quantum Center, University of the Basque Country UPV/EHU, 48940 Leioa, Bizkaia, Spain}
\email{evgeny.sherman@ehu.eus}

\begin{abstract}
Simultaneous manipulation of charge and spin density distributions in materials is the key
element required in spintronics applications. Here we study the formation of coupled spin and 
charge densities arising in scattering of electrons by domains of local magnetization 
producing a position-dependent Zeeman field
in the presence of the spin-momentum locking typical for topological insulators. 
Analytically and numerically calculated scattering pattern is determined by 
the electron energy, domain magnetization, and size. 
The spin-momentum locking produces strong differences with respect to the spin-diagonal scattering
and leads to the scattering asymmetry with nonzero mean scattering angle as determined
by only two parameters characterizing the system. To extend the variety of possible patterns, 
we study scattering by diffraction gratings and propose to design them in modern nanostructures based on topological
insulators to produce desired distributions of the charge and spin densities.  
These results can be useful for engineering of magnetic patterns for electron optics
to control coupled charge and spin evolution. 
\end{abstract}

\date{\today}
\maketitle

\section{Introduction: electron optics with spin-momentum locking}

The ability to manipulate and control electron charge and spin dynamics by external
fields is one of the challenges in modern applied physics. This goal can
be achieved by electron optics, that is using elements similar to
conventional optics based on wave properties of electrons. Another
option is related to the electron spin optics, that is to control both
coupled electron spins and charge dynamics. A conventional tool to manipulate electron
spin is the Zeeman-like coupling either with the external magnetic field or with material 
magnetization. However, simultaneous control of spin and
charge motion requires spin-orbit coupling \cite{Bychkov1984} resulting in spin-momentum locking. 
This can be achieved by using electrons in {low-energy two-dimensional surface states of topological insulators,} 
where this coupling demonstrates itself
as a strong spin-momentum locking expressed in the Hamiltonian \cite{Hasan2010,Qi2011} 
\begin{equation}\label{eq:Hamiltonian}
H=-i\hbar v\bsig\cdot {\bm\nabla +}\Delta_{\rm m} (\mathbf{r})\sigma_{z},
\end{equation}%
{and produces relativistic-like Dirac cones.} 
Here position $\mathbf{r}=(x,y),$ $v$ is the electron bandstructure velocity parameter and $\Delta_{\rm m} (\mathbf{r})$ is the
local magnetization assumed to be along the $z-$axis. We assume that the electron energy is considerably low 
{and the corresponding momentum is sufficiently small} to satisfy
the validity of the Hamiltonian \eqref{eq:Hamiltonian}, including only the linear
$\bsig\cdot{\bm\nabla}$ term and neglecting the band warping \cite{Fu2009}, with $\sigma_{i}$ being the Pauli
matrices. Here it is convenient to present the electron spin in the form 
${\bm\sigma}=({\bm\sigma }_{\perp},\sigma_{z}),$ where ${\bm\sigma}_{\perp
}=(\sigma_{x},\sigma_{y})$ is the two-dimensional in-plane component. The
velocity of the electron defined as $\mathbf{v}=i[H,\mathbf{r}]/\hbar
=v(\sigma_{x},\sigma_{y})=v{\bm\sigma}_{\perp}$ is determined by the electron spin. 
Thus, by acting at the electron spin by a position- and
time-dependent Zeeman field one can modify the electron velocity 
and, thus, influence the charge transport 
(see, e.g., Ref. [\onlinecite{Yokoyama2011}] for electron propagation in the presence of
one-dimensional stripe-like magnetization). Recently, Ref. [\onlinecite{Mochida2022}] analyzed 
the effects of skew scattering by magnetic monopoles in spin-orbit 
coupled systems.

On the one hand, random magnetic disorder modifies weak localization \cite{Hikami1980} in
conventional semiconductors and strongly influences the conductivity of topological
insulators \cite{Kudla2019}. On the other hand, the ability to produce magnetization pattern  
$\Delta_{\rm m} (\mathbf{r})$ \cite{Katmis2016,Tokura2019}  permits design of
position-dependent spin dynamics and, correspondingly, the manipulation of
electron wavefunction producing the coupled spin-charge transport {\cite%
{Jamali2015,Chiba2017,Chen2020}}. The effects of spin-dependent velocity are known for conventional
semiconductors for electrons \cite{Rokhinson2004} and holes \cite{Rendell2022} and can manifest
itself in electron scattering by impurities in the presence of strong spin-orbit 
coupling \cite{Hutchinson2017}, formation of equilibrium spin currents \cite{Rashba2004,Tokatly2010} and in 
the spin-Hall effect \cite{Dyakonov1971,Sinova2004,Murakami2003,Kato2004,Mishchenko2004,Sih2005,Wunderlich2005}
kinetics. 

Here we explore the possibility to control electron spin and position
dynamics by a single- and arrays of magnetized quantum dots with spatially confined magnetization on
the surface of topological insulators. One possible 
application is the variety of spin torques \cite{Mellnik2014,Ndiaye2017,Han2017,Ghosh2017,Mahendra2018,Moghaddam2020,Han2021,
Bondarenko2017} produced on the magnetized quantum dots by
scattered electrons to manage magnetization dynamics and coupled spin-charge transport. We study the scattering processes
in different regimes and conclude that a net charge current injection can be produced. 
Based on the results of a single-dot scattering approach, 
we propose to design diffraction gratings made by a one-dimensional array 
of magnetic quantum dots for this purpose.
This grating will produce on purpose asymmetric patterns of spin and charge densities and
currents.

The rest of the paper is organized as follows. In Sec. II we formulate the scattering problem 
and present the observables of interest.
In Sec. III we describe partial wave summation approach and present general  analytical results. 
Different sets of parameters and scattering domains for single scatterers will be analyzed 
in Sec. IV while scattering by diffraction grating will be considered in Sec. V. 
The main numerical results for cross-section and scattering angles will be presented in Sec. VI.
Section VII provides the conclusions and outlook of this paper.  

\section{Scattering process and observables}

\begin{figure}[tbp]
\includegraphics*[width=0.4\textwidth]{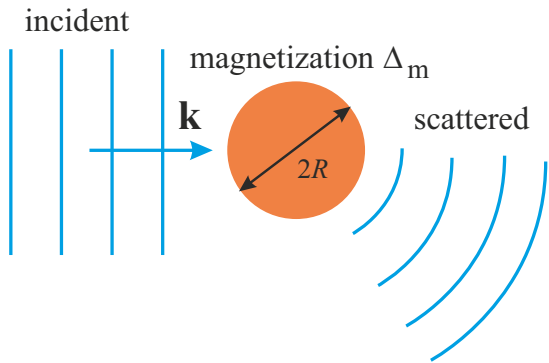}\\
\caption{Asymmetric scattering of electron with the wavevector $\mathbf{k}$ by a single magnetized nanodot.}
\label{fig:single}
\end{figure}

For a circular magnetized disk on the surface of topological insulator we rewrite Eq. \eqref{eq:Hamiltonian} in the form: 
\begin{equation}
H=-iv\bsig\cdot {\bm\nabla}+\theta(R-r)\Delta_{\rm m}\sigma_{z},
\label{eq:Hband}
\end{equation}%
where $\theta (R-r)\Delta_{\rm m}$ is the local magnetization with the Heaviside
function $\theta (R-r),$  as shown in Fig. \ref{fig:single}. Here and below we use the system of units with 
$\hbar\equiv\,1.$

In the absence of external magnetization $(\Delta_{\rm m} \equiv 0)$ the free-space
plane-wave function for ${\bm\psi }_{\mathbf{k}}(\mathbf{r})\sim e^{i\mathbf{%
k}\cdot \mathbf{r}}[\psi^{\uparrow}_{\mathbf{k}},\psi^{\downarrow}_{\mathbf{k}}]^{\mathrm{T}}$ (T
stands for transposition) satisfies the equation: 
\begin{equation}
\left[ 
\begin{array}{cc}
-\varepsilon  & vk_{-} \\ 
vk_{+} & -\varepsilon 
\end{array}%
\right] \left[ 
\begin{array}{c}
\psi^{\uparrow}_{\mathbf{k}} \\ 
\psi^{\downarrow}_{\mathbf{k}}%
\end{array}%
\right] =0,
\end{equation}%
where $k_{\pm }\equiv \,k_{x}\pm ik_{y}$. We obtain two linear branches of
spectrum $\varepsilon =\pm vk$ where $k=\sqrt{\,k_{x}^{2}+k_{y}^{2}}$ . For
the eigenstates one has%
\begin{equation}
{\bm\psi }_{\mathbf{k}}(\mathbf{r})=\frac{e^{i\mathbf{k}\cdot \mathbf{r}}}{%
\sqrt{2}}\left[ 
\begin{array}{c}
1 \\ 
\varepsilon\,k_{+}/|\varepsilon|\,k%
\end{array}%
\right] .
\end{equation}
and the spin is parallel $\left(\varepsilon >0\right) $ or antiparallel $%
\left(\varepsilon <0\right) $ to the momentum. At a large distance from the
quantum dot the electron wavefunction at $\varepsilon >0$ is presented for
the wave coming from $x=-\infty $ along the $x-$axis with $\mathbf{k}=(k,0)$ [\onlinecite{Newton2013}]
\begin{equation}
{\bm\psi }(r,\varphi )=\frac{e^{ikx}}{\sqrt{2}}\left[ 
\begin{array}{c}
1 \\ 
1%
\end{array}%
\right] +{\bm f}(\varphi )\,\frac{e^{ikr}}{\sqrt{r}},
\end{equation}%
where 
\begin{equation}
{\bm f}(\varphi )=\left[ 
\begin{array}{c}
f^{\uparrow }(\varphi ) \\ 
f^{\downarrow}(\varphi )%
\end{array}%
\right] 
\end{equation}%
is the two-component spinor scattering amplitude. We note that in two-dimensional systems the  
cross-section $l$ has the length units and present the differential
cross-section as $dl/d\varphi =|{\bm f}(\varphi )|^{2}$ where the total $l$ is: 
\begin{equation}
l=\int_{-\pi }^{\pi }|{\bm f}(\varphi )|^{2}d\varphi .
\label{eq:ltot}
\end{equation}%
Since, as we will demonstrate below, the scattering is anisotropic with $%
|{\bm f}(-\varphi )|\neq|{\bm f}(\varphi)|,$ we introduce the mean 
value of the scattering angle $\langle \varphi\rangle$ and of its square $\left\langle \varphi
^{2}\right\rangle:$ 

\begin{equation}\label{eq:varphi_n}
\langle \varphi^{n}\rangle =\frac{1}{l}\int_{-\pi}^{\pi }\varphi^{n}
|{\bm f}(\varphi)|^{2}d\varphi,
\end{equation}%
where $n=1$ or $n=2.$
The dispersion $D_{\varphi}=\sqrt{\langle\varphi^{2}\rangle -\langle\varphi\rangle^{2}}$, 
characterizes the width of the scattering aperture.

In addition, we mention that the asymmetric scattering produces effective
charge current along the $y-$axis, which can be defined as:
\begin{equation}
\langle j\rangle = \frac{ev}{l}\int_{-\pi}^{\pi}\sin\varphi |{\bm f}(\varphi)|^{2}d\varphi 
\end{equation}%
where $e$ is the electron charge. The behavior of this current as a function of system
parameters is qualitatively similar to 
the behavior of $\left\langle\varphi\right\rangle.$

\section{Partial wave summation: analytical results}

\subsection{Wave functions and boundary conditions}

In polar coordinates with $r=\sqrt{x^{2}+y^{2}}$, $\varphi =\arctan(y/x)$,
the eigenstates in the form of the circular waves are determined by: 
\begin{equation}
\left[ 
\begin{array}{cc}
\varepsilon -\Delta_{\rm m} (r) & ve^{-i\varphi }\left(i\,\partial_{r}+%
\displaystyle{\frac{\partial_{\varphi }}{r}}\right) \\ 
ve^{i\varphi }\left(i\,\partial_{r}-\displaystyle{\frac{\partial_{\varphi
}}{r}}\right) & \varepsilon +\Delta_{\rm m} (r)%
\end{array}%
\right] {\bm\psi}(r,\varphi)=0.  \label{eq:SE1}
\end{equation}%
{To calculate the sum of partial waves attributed to the $z-$components of the
angular momentum $m,$ we first substitute in Eq. \eqref{eq:SE1} the spinor
characterized by given $m$ in the form:} 
\begin{equation}\label{spinor}
{\bm\psi }_{m}(r,\varphi )=\,e^{im\varphi }\left[ 
\begin{array}{c}
\psi^{\uparrow}_{m}(r) \\ 
\psi^{\downarrow}_{m}(r)\,e^{i\varphi }%
\end{array}%
\right]
\end{equation}%
{and obtain} coupled equations for the radial functions (omitting the
explicit $r-$dependence for brevity): 
\begin{equation}
\left[ 
\begin{array}{cc}
\Delta_{\rm m} (r)-\varepsilon & -iv\left(\displaystyle{\frac{d}{dr}+\frac{m+1}{r}}\right) \\ 
-iv\left(\displaystyle{\frac{d}{dr}-\frac{m}{r}}\right) & -\Delta_{\rm m} (r)-\varepsilon%
\end{array}%
\right] \left[ 
\begin{array}{c}
\psi^{\uparrow}_{m}\\ 
\\
\\
\psi^{\downarrow}_{m}%
\end{array}%
\right] =0.  \label{eq:SEC}
\end{equation}

Inside the dot, $r<R$ and $\Delta_{\rm m} (r)=\Delta_{\rm m} :$ 
\begin{eqnarray}
&&(\Delta_{\rm m} \,-\varepsilon )\,\psi^{\uparrow}_{m}-
iv\left({\frac{d}{dr}+\frac{m+1}{r}}\right)\psi^{\downarrow}_{m}=0 \label{eq:SER1} \\
&&-iv\left({\frac{d}{dr}-\frac{m}{r}}\right)\psi^{\uparrow}_{m}-(\Delta_{\rm m} \,+\varepsilon
)\,\psi^{\downarrow}_{m}=0.  \label{eq:SER2}
\end{eqnarray}%
Extracting $\psi^{\downarrow}_{m}$ in Eq. (\ref{eq:SER2}) 
\begin{equation}
\psi^{\downarrow}_{m}=-\frac{iv}{\Delta_{\rm m} \,+\varepsilon }\left({\frac{d}{dr}-\frac{m}{r}}\right)\psi^{\uparrow}_{m}   \label{eq:chim}
\end{equation}%
and substituting it into (\ref{eq:SER1}) we obtain for $\psi^{\uparrow}_{m}$ 
\begin{equation}
\kappa ^{2}\psi^{\uparrow}_{m}-\left(\frac{d^{2}}{dr^{2}}+\frac{1}{r}\frac{d}{dr}-\frac{m^{2}}{r^{2}}\right)\psi^{\uparrow}_{m} =0,
\end{equation}
with the energy-dependent $\kappa\equiv\sqrt{|\Delta_{\rm m}^{2}-\varepsilon^{2}|}/v.$ 
We begin with the realization $\varepsilon<\Delta_{\rm m}$
where the solution regular at $r\rightarrow 0$ is the modified Bessel
function $\psi^{\uparrow}_{m}(r\kappa)=I_{m}(r\kappa)$ [\onlinecite{Lebedev}]  and
then consider the $\varepsilon>\Delta_{\rm m}$ case by analytical continuation. Using (\ref%
{eq:chim}) we can write: 
\begin{equation}
\psi^{\downarrow}_{m}(z)=-is\left(\frac{d}{dz}-\frac{m}{z}\right)\psi^{\uparrow}_{m}(z)
=-isI_{m+1}(z),
\end{equation}%
where $z\equiv r\kappa$ and $s\equiv\sqrt{|\Delta_{\rm m}-\varepsilon|/(\Delta_{\rm m}+\varepsilon)}.$

Thus, the general solution at $r<R$ is 
\begin{equation}
\left[ 
\begin{array}{c}
\psi^{\uparrow}_{m} \\ 
\psi^{\downarrow}_{m}%
\end{array}%
\right] =A_{m}\left[ 
\begin{array}{c}
I_{m}(\kappa r) \\ 
-isI_{m+1}(\kappa r)%
\end{array}%
\right] ,  \label{eq:inside}
\end{equation}%
where $A_{m}$ is a constant.

For $r>R$ with $\Delta_{\rm m}=0$ introducing for brevity $\mu\equiv\,r\varepsilon /v=kr$ we obtain 
\begin{equation}
\left[\mu^{2}\frac{d^2}{d\mu^{2}}+\mu\frac{d}{d\mu}+(\mu^{2}-m^{2})\right]
\psi^{\uparrow}_{m}=0,
\end{equation}%
with the Bessel functions $J_{m}(kr)$ and $Y_{m}(kr)$ [\onlinecite{Lebedev}]
solutions where $\psi^{\downarrow}_{m}$ being expressed with $J_{m+1}(kr)$ and $%
Y_{m+1}(kr).$  The resulting general solution at $r>R$ is the
superposition of the waves with harmonics 
\begin{equation}
\left[ 
\begin{array}{c}
\psi^{\uparrow}_{m} \\ 
\psi^{\downarrow}_{m}%
\end{array}%
\right] =B_{m}\left[ 
\begin{array}{c}
J_{m}(kr) \\ 
iJ_{m+1}(kr)%
\end{array}%
\right] +C_{m}\left[ 
\begin{array}{c}
Y_{m}(kr) \\ 
iY_{m+1}(kr)%
\end{array}%
\right] .
\end{equation}

These equations are supplemented by the two continuity conditions at $r=R$ which
can be reduced to a single equation as: 
\begin{eqnarray}
&& B_{m}J_{m}(kR)+C_{m}Y_{m}(kR) \label{eq:cont2} \\
&=&-\frac{I_{m}(\kappa R)}{sI_{m+1}(\kappa R)}\left(
B_{m}J_{m+1}(kR)+C_{m}Y_{m+1}(kR)\right). \notag 
\end{eqnarray}

\subsection{Scattering amplitude: summed partial waves}

To perform summation over harmonics $m,$ we begin with the plane wave
resolution $e^{ikr\cos \varphi }=\sum_{m=-\infty }^{\infty
}i^{m}e^{im\varphi }\,J_{m}(kr)$\cite{smythe} resulting in 
\begin{equation}
{\bm\psi }_{\mathbf{k}}(r,\varphi )=\frac{1}{\sqrt{2}}\left[ 
\begin{array}{c}
1 \\ 
1%
\end{array}%
\right] \sum_{m=-\infty }^{\infty }i^{m}e^{im\varphi }\,J_{m}(kr).
\end{equation}%
At large distances, $kr\gg |m^{2}-1/4|$, we use asymptotics of the Bessel
functions \cite{Lebedev}: 
\begin{eqnarray} \label{asymptoticsJ}
{J}_{m}\left(kr\right) &\simeq &\sqrt{\frac{2}{\pi kr}}
\cos\left(kr-\frac{m\pi }{2}-\frac{\pi}{4}\right), \\
Y_{m}\left(kr\right) &\simeq &\sqrt{\frac{2}{\pi kr}}
\sin\left(kr-\frac{m\pi}{2}-\frac{\pi}{4}\right). \label{asymptoticsY}
\end{eqnarray}

By using condition that the wave function contains only the outgoing $\exp
(ikr)$ and no ingoing $\exp(-ikr)$ waves \cite{Newton2013}, that is the ingoing wave terms 
mutually cancel each other, we obtain 
\begin{eqnarray}
f^{\uparrow }(\varphi )&=&-ie^{-i\pi/4}\sqrt{\frac{1}{\pi k}}\sum_{-\infty
}^{\infty }\frac{\gamma_{m}}{1+i\gamma_{m}}e^{im\varphi},  \label{eq:fup}
\\
f^{\downarrow}(\varphi)&=&f^{\uparrow}\left(\varphi\right)e^{i\varphi}.  \label{eq:fdown}
\end{eqnarray}%
Here $\gamma_{m}\equiv \,C_{m}/B_{m}$ is obtained with the boundary
conditions in the form of Eq. \eqref{eq:cont2}. The spin component expectation values for the
scattered wave defined as 
\begin{equation}
\sigma_{i}(\varphi )=\left\langle{\bm f}(\varphi)|\sigma_{i}|{\bm f}(\varphi)\right\rangle 
\end{equation}%
are the same as those of the free state since $k_{+}=ke^{i\varphi}$
with $\sigma_{z}(\varphi)=0.$

In the low-energy domain $\varepsilon<\Delta_{\rm m} $ we obtain 
\begin{equation}
\gamma_{m}=-\frac{sJ_{m}(kR)I_{m+1}(\kappa R)+I_{m}(\kappa R)J_{m+1}(kR)}{%
sY_{m}(kR)I_{m+1}(\kappa R)+I_{m}(\kappa R)Y_{m+1}(kR)}.  
\label{eq:gammam}
\end{equation}

Similar calculation for the high-energy domain $\varepsilon >\Delta_{\rm m} $  using 
relation $I_{m}(i|z|)=\left(
-i\right) ^{m}J_{m}\left(-|z|\right)$ and $J_{m}(-x)=(-1)^{m}J_{m}(x)$ 
yields

\begin{equation}\label{eq:gammamlarge}
\gamma_{m}=-\frac{sJ_{m}(kR)J_{m+1}(\kappa R)-J_{m}(\kappa R)J_{m+1}(kR)}{%
sY_{m}(kR)J_{m+1}(\kappa R)-J_{m}(\kappa R)Y_{m+1}(kR)}.
\end{equation}%
Various scattering regimes described by these equations will be discussed below 
analytically and numerically.

\subsection{Scattering cross-section and asymmetry}

It is convenient to introduce even $\Gamma_{g}\left(
m_{1},m_{2}\right) =\Gamma_{g}\left(m_{2},m_{1}\right) $ and odd $\Gamma
_{u}\left(m_{1},m_{2}\right) =-\Gamma_{u}\left(m_{2},m_{1}\right) $
matrices as
\begin{eqnarray}
&&\frac{\gamma_{m_{1}}}{1-i\gamma_{m_{1}}}\frac{\gamma_{m_{2}}}{1+i\gamma_{m_{2}}}
=\Gamma_{g}\left(m_{1},m_{2}\right) +\Gamma_{u}\left(m_{1},m_{2}\right) \label{eq:Gamma}\\
&&\Gamma_{g}\left(m_{1},m_{2}\right) = \frac{\gamma_{m_{1}}\gamma_{m_{2}}}{%
\left(1+\gamma_{m_{1}}^{2}\right) \left(1+\gamma_{m_{2}}^{2}\right) }\left(
1+\gamma_{m_{1}}\gamma_{m_{2}}\right) \nonumber \\
&&\Gamma_{u}\left(m_{1},m_{2}\right) = i\frac{\gamma_{m_{1}}\gamma_{m_{2}}}{%
\left(1+\gamma_{m_{1}}^{2}\right) \left(1+\gamma_{m_{2}}^{2}\right) }\left(
\gamma_{m_{1}}-\gamma_{m_{2}}\right) \nonumber,
\end{eqnarray}%
and use them to define $l/R$, $\left\langle \varphi \right\rangle $ and $%
\left\langle \varphi ^{2}\right\rangle.$ 
Using Eqs. \eqref{eq:varphi_n}, \eqref{eq:fup}, \eqref{eq:fdown}, and \eqref{eq:Gamma},
the $l/R$ and the mean value $\langle\varphi\rangle$ can be expressed with these matrices as:
\begin{eqnarray}
\frac{l}{R}&=&\frac{4}{kR}\sum_{m}\frac{\gamma_{m}^{2}}{1+\gamma_{m}^{2}}=\frac{4}{kR}{\rm tr}\,\Gamma_{g}\left(m_{1},m_{2}\right),\\ 
\langle\varphi\rangle &=& -\frac{4i}{kl}\sum_{m_{1},m_{2}} 
\Gamma_{u}\left(m_{1},m_{2}\right)
\frac{\left(-1\right)^{m_{2}-m_{1}}}{m_{2}-m_{1}}
\end{eqnarray}%
making the scattering asymmetric with nonzero $\langle\varphi\rangle$ 
solely due to the imaginary terms $i\gamma_{m}$ in the denominators of Eqs. \eqref{eq:fup}, \eqref{eq:fdown},  which appear
due to the phase shift between the spin components in Eq. (\ref{spinor}). This effect is qualitatively
different from the spin-diagonal scattering by a radially-symmetric 
potential, which is always $\varphi\leftrightarrow\,-\varphi$ symmetric, and is similar to the scattering mechanisms producing 
the anomalous Hall effect \cite{Nagaosa2010}.
For the $\langle\varphi^{2}\rangle$ we obtain similarly:%
\begin{equation}
\langle\varphi^{2}\rangle = \frac{8}{kl}\sum_{m_{1},m_{2}}
\Gamma_{g}\left(m_{1},m_{2}\right)\frac{\left(-1\right)^{m_{2}-m_{1}}}{\left(m_{2}-m_{1}\right)^{2}}%
+\frac{\pi^{2}}{3}.
\end{equation}%

\section{Sets of parameters and scattering domains}

We introduce two parameters which fully describe the scattering process $%
M\equiv \,R\Delta_{\rm m} /v,\epsilon \equiv kv/\Delta_{\rm m} $ and express the scattering
amplitudes with

\begin{equation}
s=\sqrt{\frac{|1-\epsilon |}{1+\epsilon }};\qquad
\kappa R=M\sqrt{|1-\epsilon ^{2}|};\qquad kR=M\epsilon,
\end{equation}%
where $\kappa /k=\sqrt{|1-\epsilon ^{2}|}/\epsilon .$
Parameter $M$ corresponds to the angular momentum of the
electrons with the resonant energy $\Delta_{\rm m}$ and can be seen as $\tau_{p}\Delta_{\rm m},$
where $\tau_{p}=R/v$ is the typical passing time through the magnetic domain while the limit
$M\ll\,1$ corresponds to the Born approximation of the scattering theory \cite{Kudla2022}. 

\subsection{Low-energy domain $\epsilon <1$}

We consider first the low-energy domain $\epsilon <1.$ To demonstrate the
main properties of the scattering, we begin with the small-radius, large
wavelength limit $kR\ll 1,$ where spin-independent scattering theory predicts
angle-independent probability with $|{\bm f}(\varphi)|^{2}=\mathrm{const}.$
As we will show, however, it is not the case in the presence of
spin-momentum locking. For this purpose we use small-$x$ behavior of the
Bessel functions:%
\begin{eqnarray}
J_{m}(x) &\simeq &I_{m}(x)\simeq \frac{1}{m!}\left(\frac{x}{2}\right)
^{m}; \\
Y_{m}(x) &\simeq &-\frac{(m-1)!}{\pi }\left(\frac{x}{2}\right) ^{-m} \nonumber
\end{eqnarray}%
and their index-parity transformations:%
\begin{eqnarray}
J_{-m}(x) &=&\left(-1\right) ^{m}J_{m}(x);\quad I_{-m}(x) =I_{m}(x), \\
Y_{-m}(x) &=& \left(-1\right)^{m}Y_{m}(x). \nonumber
\end{eqnarray}

 We consider first a nonresonant scattering with $kR\ll\,1$ and $%
\kappa R\gg\,kR.$ Thus, we select terms by the
lowest powers of $kR$ in the numerator \cite{Newton2013,Landau1981} and highest powers of $\left(
kR\right) ^{-1}$ in the dominator and obtain for $m\geq 0$ 
\begin{eqnarray}
\gamma_{m\geq 0}&=&-\frac{J_{m}(kR)I_{m+1}(\kappa R)}{I_{m}(\kappa
R)Y_{m+1}(kR)} \\
&=&\frac{I_{m+1}(\kappa R)}{I_{m}(\kappa R)}\frac{\pi }{(m!)^{2}}
\left(\frac{kR}{2}\right)^{2m+1}.\notag
\end{eqnarray}%
Making similar $kR-$powers selection for $m<0$ we obtain: 
\begin{equation}
\gamma_{m<0}=-\frac{I_{m}(\kappa R)}{I_{m+1}(\kappa R)}\frac{\pi }{\left(|m+1|!\right)^{2}}
\left(\frac{kR}{2}\right)^{2|m|-1}
\end{equation}%
and see fast decrease with $|m|$ both for positive and
negative $m.$ Therefore, at $kR\ll 1$ and $M\sim \,1$ one obtains the
resulting angular dependence $|{\bm f}(\varphi )|^{2}\sim \sin ^{2}\varphi/2$ 
with predominant backscattering, qualitatively different from the spin-diagonal 
scattering \cite{Landau1981}. The ratio $l/R\sim kR$ is linear in the
energy and a weak asymmetry $\left\langle \varphi
\right\rangle\sim \gamma_{0}-\gamma_{-1}\sim kR\sim\, l/R$ \cite{Kudla2022}. In this limit $%
\left\langle \varphi ^{2}\right\rangle=2+\pi ^{2}/3$ and, therefore, $%
D_{\varphi }=\sqrt{2+\pi ^{2}/3}.$

Next, we turn to small wavelength, large radius limit $kR\gg |m^{2}-1/4|$ away from resonance with
$\kappa R\gtrsim\,1$ but $\kappa R<|m^{2}-1/4|.$ Then, by using asymptotics for the 
functions of $kR$ and exact expressions for the functions of $\kappa R$ and noticing that 
no power selection is required here, we obtain after a straightforward calculation:%
\begin{equation}
\gamma_{m}=-\tan\left(kR-\frac{\pi m}{2}-\frac{\pi}{4}+\xi_{m}\right)
\end{equation}%
with $\xi_{m}=\arctan\left(sI_{m+1}(\kappa R)/I_{m}(\kappa R)\right).$

\subsection{Resonant scattering $\epsilon \rightarrow 1$}

Next, consider resonant scattering as the energy of 
electron is close to $\Delta_{\rm m} $ with $kR=M$ and $\kappa R\ll\,kR$ 
at $\epsilon \rightarrow 1.$ 
Here $s=\sqrt{(1-\epsilon)/(1+\epsilon)}\approx \sqrt{(1-\epsilon)/2}$ 
and $\kappa R=M\sqrt{2}\sqrt{1-\epsilon}$ yield $s\approx \kappa R/2M\ll 1.$ Making expansions 
in Eq. \eqref{eq:gammam}, we obtain
\begin{equation}
\gamma_{m} =-\frac{{\kappa R}J_{m}(M)I_{m+1}(\kappa
R)+{2M}I_{m}(\kappa R)J_{m+1}(M)}{{\kappa R}Y_{m}(M)I_{m+1}(\kappa R)+{2M}I_{m}(\kappa R)Y_{m+1}(M)}. 
\end{equation}

Here we perform selection by power counting of small $\kappa R$ and obtain
\begin{equation}
\gamma_{m\ge\,0} =-\frac{J_{m+1}(M)}{Y_{m+1}(M)}, 
\end{equation}
in the limit $M\ll 1$ this yields:
\begin{equation}
\gamma_{m\ge\,0}\approx\frac{\pi}{(m+1)!m!}
\left(\frac{M}{2}\right)^{2(m+1)}.
\end{equation}

For $m<0$ we take into account that: 
\begin{equation}
I_{m+1}(\kappa R)\approx\frac{1}{|m+1|!}\left(\frac{\kappa R}{2}%
\right) ^{|m|-1} 
\end{equation}%
and obtain:
\begin{equation}
\gamma_{m<0} = -\frac{|m|J_{m}(M)+MJ_{m+1}(M)}{|m|Y_{m}(M)+MY_{m+1}(M)}, 
\end{equation}%
yielding in the limit $M\ll\,1:$
\begin{equation}
\gamma_{m<0} = \frac{(-1)^{m}\pi}{|m|!(|m|-1)!}
\left(\frac{M}{2}\right)^{2|m|}.
\end{equation}
Since in this limit $\gamma_{0}=-\gamma_{-1}$ with $|\gamma_{|m|>1}|\ll |\gamma_{0}|,$ the scattering behavior 
remains the same as in the $\epsilon\ll\,1$ case.

\subsection{High-energy domain $\epsilon>1$}

We use in Eq. (\ref{eq:gammamlarge}) known asymptotics in Eqs. (\ref{asymptoticsJ}) and (\ref{asymptoticsY}) for the 
realization $kR\approx \kappa R \gg 1$ and 
where both the effective angular momentum and energy are
large. Summing the terms and taking into account that: 
$\kappa R\pm\,kR = M\left(\sqrt{\epsilon ^{2}-1}\pm\epsilon \right),$ 
we obtain for the realization $kR\approx \kappa R \gg |m^{2}-1/4|$ :
\begin{equation}
\gamma_{m} = -\frac
{(-1)^{m}(1-s)\cos\left(M\epsilon_{+}\right)
+(1+s)\sin \left(M\epsilon_{-}\right)}
{(1+s)\cos\left(M\epsilon_{-}\right)
+(-1)^{m}(1-s)\sin\left(M\epsilon_{+}\right)},
\end{equation} 
where $\epsilon_{\pm}\equiv\,\sqrt{\epsilon^{2}-1}\pm\,\epsilon.$

In the case $\epsilon \gg 1$ the leading terms in the expansion by $1/\epsilon$ result in:%
\begin{equation}\label{eq:henergy}
\gamma_{m}=(-1)^{m}\frac{\cos(2M\epsilon)}{2\epsilon}
\end{equation}
rapidly decreasing with the energy. The limit $\epsilon\rightarrow\infty$ 
evidently yields $\gamma_{m}=0.$   

\section{Scattering by diffraction grating}

\begin{figure}[tbp]
\includegraphics*[width=0.4\textwidth]{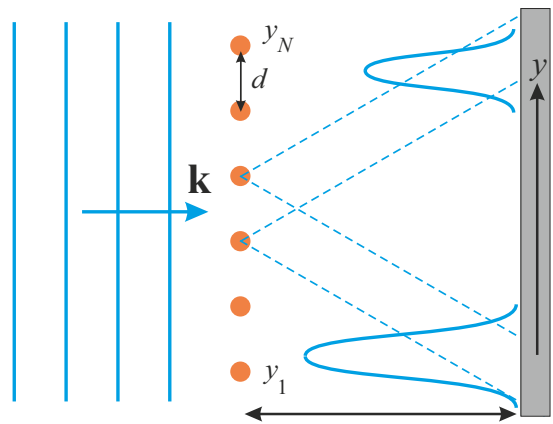}\\
\caption{Schematic plot of electron scattering by a diffraction grating. 
Only the principal peaks are shown. The presented asymmetry corresponds 
to asymmetric scattering in Fig. \ref{fig:single}.}
\label{fig:gratings}
\end{figure}

We consider now a diffraction grating formed by the linear chain 
of magnetic dots (nanodiscs) at the surface of topological insulator as 
shown schematically in Fig. \ref{fig:gratings}.  The array contains $N$ identical dot scatterers separated
by the distance $d$ such that position of the center of domain is given by: $y_{i}
=d(1-N)/2+d(i-1),$ $i=1,\ldots,N.$

In this geometry, we assume that each dot 
is an independent scatterer of the incoming plane wave with 
spin polarization along axis $x$. The distance $d$ between  
neighboring dots  is of the order of electron wavelength 
$\lambda$, and we are observing the diffraction pattern at a relatively 
large distance $L\gg \lambda $. To consider the grating as a 
chain of independent scatterers, 
we first formulate the scattering independence condition:%
\begin{equation}
\frac{R}{d}\left|{\bm f}\left({\pi}/{2}\right)\right|^{2}\ll \frac{l}{4}
\end{equation}
meaning that the wave scattered by one dot cannot be re-scattered by its neighbors.

The scattering pattern produced at points $\left(L,y\right)$ on the screen is
given by \cite{BornWolf}:
\begin{equation}
\mathbf{F}{(y)}=\sum_{i=1}^{N}{\bm f}(\varphi_{i})\,\frac{e^{ikr_{i}}}{%
\sqrt{r_{i}}},
\end{equation}%
where $r_{i}=\sqrt{\left(y-y_{i}\right)^{2}+L^{2}}$ and $\varphi
_{i}=\arctan(\left(y-y_{i}\right)/L).$ Then, we obtain the scattering
density $\left\vert\mathbf{F}{(y)}\right\vert^{2}$ and density of spin
components $\sigma_{j}(y)=\mathbf{F}^{\dag}(y)\sigma_{j}\mathbf{F}{(y)}.$

The points, where the scattered waves produce constructive interference are 
determined by the constructive interference condition, $d\sin\varphi=n\lambda.$ 
For asymmetric scattering this relation also determines the spin orientation 
in the diffraction spots. The whole diffraction picture is asymmetric as 
the "brightness" of spots is more pronounced in one of $\varphi $-directions 
(this is shown in Fig.~\ref{fig:grating} as a larger peak for $\varphi <0$ than for $\varphi >0$, 
in accordance with Fig.~\ref{fig:single}). Thus, we obtain scattering
profile corresponding to $|{f}\left(\varphi\right)|^{2}$ with asymmetric scattering pattern. 
This asymmetric profile corresponds to formation of the spin current also.
      
Now we turn to the diffraction picture for the spin polarization where the qualitative feature 
is the emergence of nonzero $z-$axis spin polarization. To clarify this effect 
we consider two-dots realization with $N=2,$  $y_{1}=-d/2,$ and $y_{2}=d/2,$ where 
\begin{equation}\label{eq:Fy2}
\mathbf{F}{(y)}={\bm f}(\varphi_{1})\,\frac{e^{ikr_{1}}}{\sqrt{r_{1}}}+{\bm %
f}(\varphi_{2})\,\frac{e^{ikr_{2}}}{\sqrt{r_{2}}},
\end{equation}
with
\begin{eqnarray}
r_{1,2}&=&\overline{r}+\Delta r_{1,2}=\overline{r}\pm\frac{d}{2}\frac{y}{\sqrt{L^{2}+y^{2}}}, \\
\varphi_{1,2}&=&\overline{\varphi} +\Delta\varphi_{1,2}=\overline{\varphi}\pm\frac{d}{2}\frac{L}{L^{2}+y^{2}}, 
\end{eqnarray}
where $\overline{r}=\sqrt{L^{2}+y^{2}}$ and $\overline{\varphi}=\arctan(y/L).$ Expansions in Eq. \eqref{eq:Fy2} 
with small $|\Delta r|\ll L$ and $|\Delta\varphi|\ll|\overline{\varphi}|\sim 1,$ show for the scattering density:
\begin{equation}\label{eq:FF}
\left\vert \mathbf{F}{(y)}\right\vert ^{2}
\approx\frac{4}{\overline{r}}\cos^{2}\left(k\Delta r\right)
\left\vert{\bm f}(\overline{\varphi})\right\vert^{2} 
\end{equation}
and for the $z-$component of spin: 
\begin{eqnarray}\label{eq:Fsigmaz}
&&\mathbf{F}^{\dag}(y)\sigma_{z}\mathbf{F}{(y)} =
\left\vert {F}^{\uparrow}(y)\right\vert ^{2}-\left\vert {F}^{\downarrow}(y)\right\vert ^{2} \\
&\approx &-\frac{1}{\overline{r}}
\sin\left(k\Delta r\right)\Delta\varphi\vert{\bm f}(\overline{\varphi})\vert^{2}, \notag 
\end{eqnarray}
where  $\Delta{r}=\Delta{r}_{2}-\Delta{r}_{1}$ and
$\Delta\varphi=\Delta\varphi_{2}-\Delta\varphi_{1}.$
Since $k\Delta r=2\pi ({d}/{\lambda})\times{y}/\sqrt{L^{2}+y^{2}}$ at $d\ll\,L$ is much larger than 
$\Delta \varphi,$ the scattering intensity weakly depends on $\Delta\varphi.$
The resulting $\mathbf{F}^{\dag}(y)\sigma_{z}\mathbf{F}{(y)}$
is not zero but rapidly decreases with $y.$

\section{Numerical results: cross-section and scattering angles}

\subsection{Single scatterers}

\begin{figure}[h]
\includegraphics*[width=0.4\textwidth]{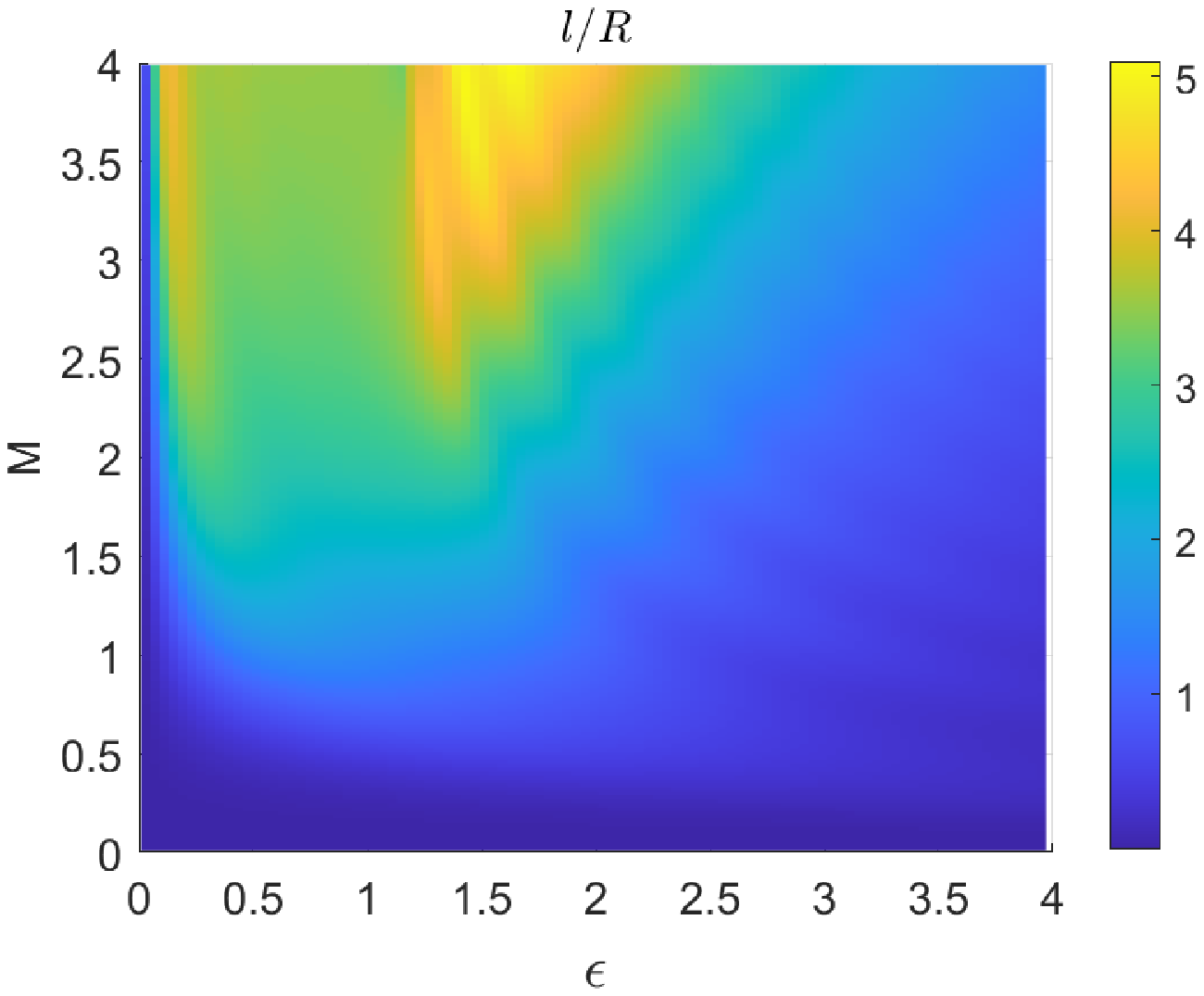}\\
\includegraphics*[width=0.4\textwidth]{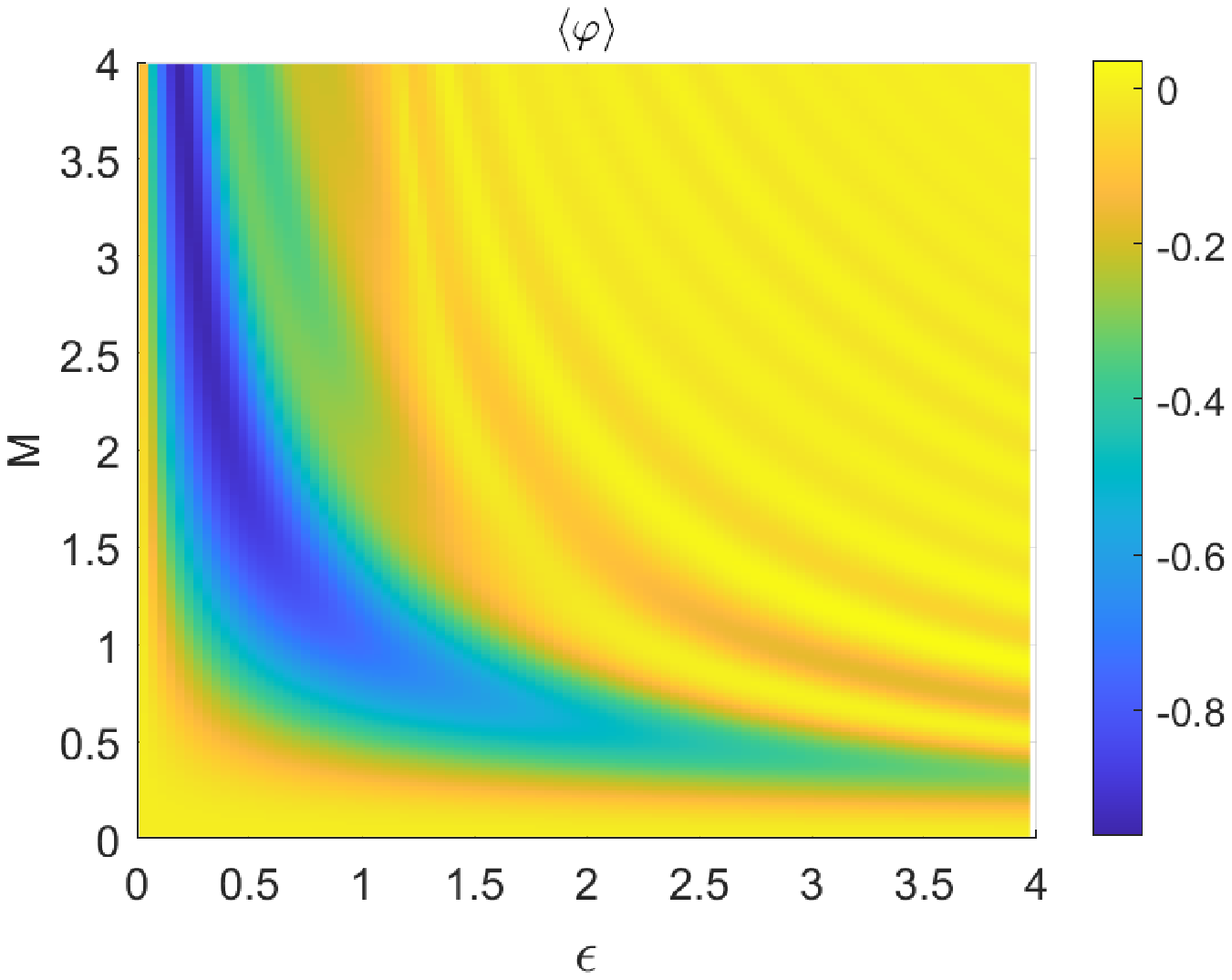}\\
\includegraphics*[width=0.4\textwidth]{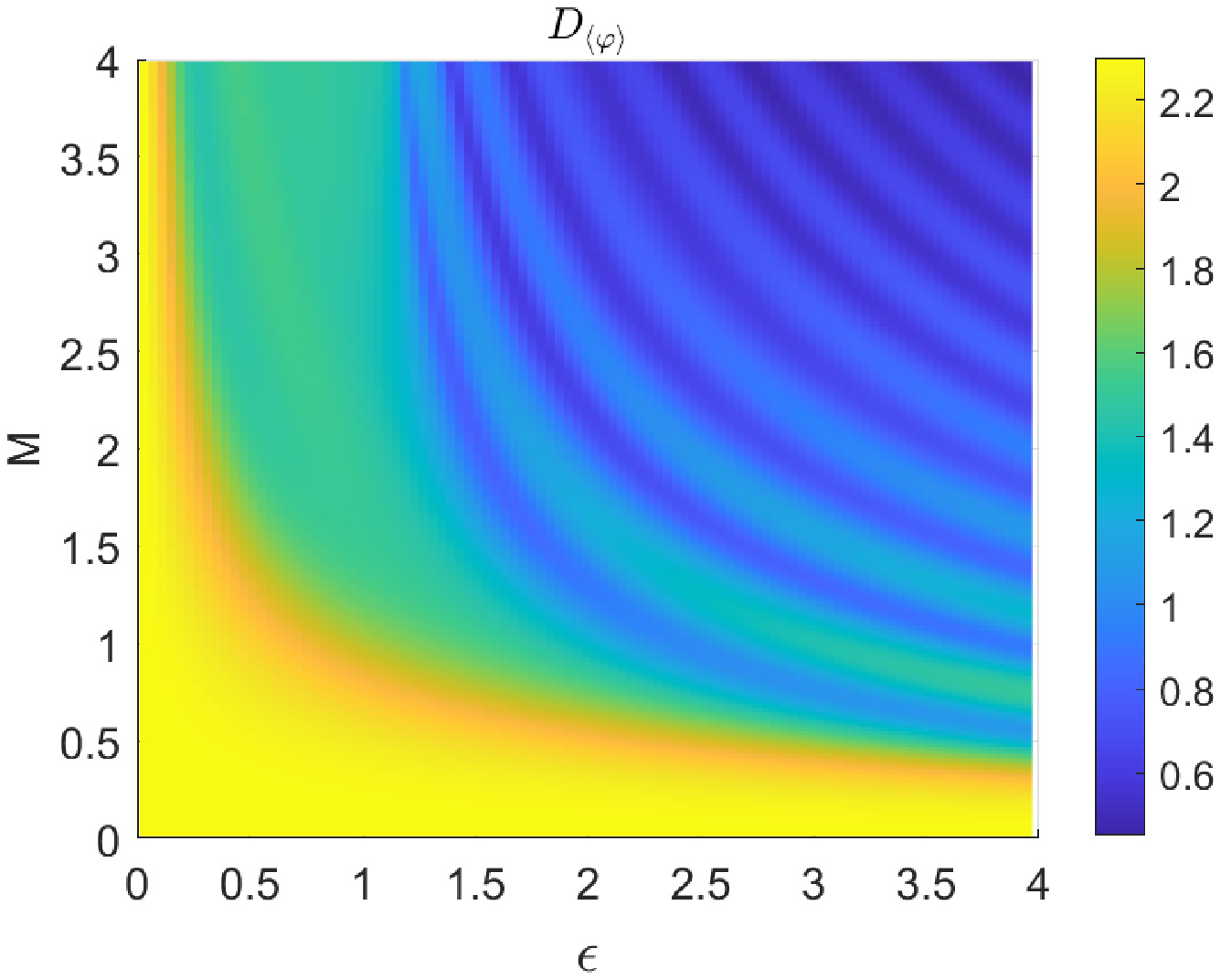}\\
\caption{Numerical results for $l/R$, $\langle\varphi\rangle,$ and $D_{\varphi},$ as marked above the plots.
{In all numerical calculations we use  $\hbar v= 2.5\times 10^{2}$ meVnm with $v=4\times 10^{7}$ cm/s typical for 
Bi-based topological insulators \cite{Liu2010}, 
$\Delta_{\rm m}=25$ meV and consider $-10\le m\le 10$ harmonics. We emphasize that the results in terms of $M$ and $\epsilon$ parameters 
are universal and do not depend on the choice of these numerical values.}}
\label{fig:3Dplots}
\end{figure}

The results of numerical calculations of the scattering cross-section length  $l/R,$ 
mean angle $\langle\varphi\rangle$ and dispersion $D_{\varphi}$
based on Eqs. \eqref{eq:gammam} and \eqref{eq:gammamlarge} are presented in Fig. \ref{fig:3Dplots} as the universal functions 
of parameters $\epsilon$ and $M$. 

The upper panel shows that the ratio $l/R$ is small both for small $M,$ corresponding to the Born approximation 
and for relatively large $M$ and $\epsilon,$  where electron energy is sufficient to ensure a relatively 
weak effect of the nanosize dot on the electron propagation. 

The mean scattering angle in the middle panel
is typically small since the scattering is still close to $\varphi\leftrightarrow\,-\varphi$
symmetric in the domain of $\epsilon\sim 1$ and
$M>1/2.$ Also, at large energies the forward scattering dominates leading to a small mean 
$|\langle\varphi\rangle|\ll 1.$

On the contrary, $D_{\varphi}$ is relatively large at $M<1$ being of the order of one
and then decreases since the forward scattering with $|\langle\varphi\rangle|\ll\,1$ and 
$\langle\varphi^{2}\rangle\ll\,1$  becomes dominating.  Notice hyperbolic structure clearly seen at $M\epsilon>1$ in
the mean scattering angle and its dispersion demonstrating 
a periodic dependence on $M\epsilon$ product  with the $\pi/2$ period,
corresponding to Eq. (\ref{eq:henergy}). 

\begin{figure}[tbp]
\includegraphics*[width=0.4\textwidth]{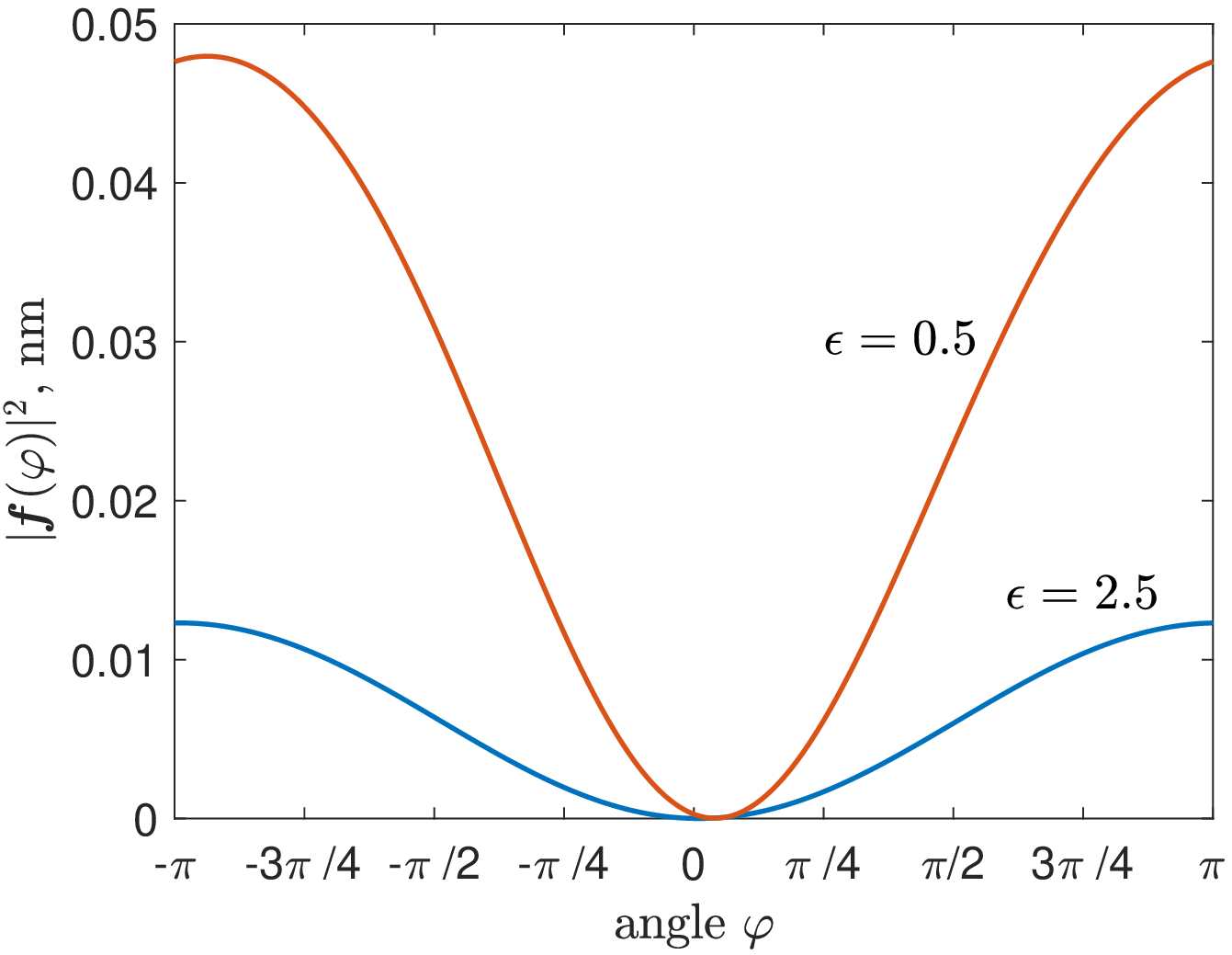}\\
\includegraphics*[width=0.4\textwidth]{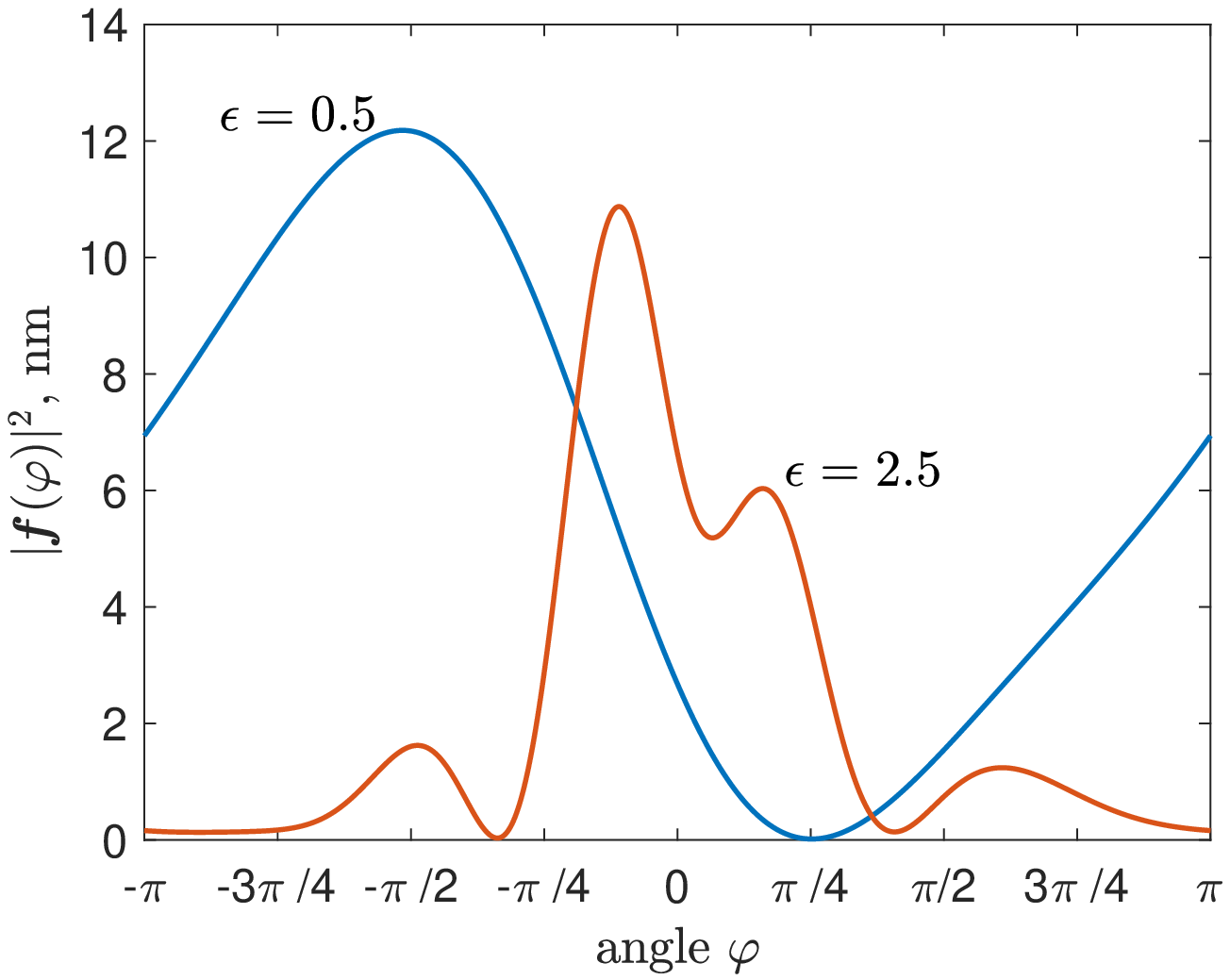}\\
\caption{Differential cross-section length of electron scattering of
magnetic quantum dot for different electron energies (as marked near the
plots). Upper panel corresponds to $R=2$ nm ($M=0.2$) and lower panel corresponds to $R=15$ nm ($M=1.5$). 
The system parameters are the same as in Fig. \ref{fig:3Dplots}. Weak forward scattering at $M\ll\,1$ 
corresponds to the perturbation theory in the Born approximation \cite{Kudla2022}.}
\label{fig:1Dexamples}
\end{figure}

To illustrate this behavior of the cross-section and scattering angle, 
we plot in Fig. \ref{fig:1Dexamples} the angular dependence of the differential scattering
cross-section. Figure \ref{fig:1Dexamples} shows that at small energies this
function behaves as $\sin^{2}(\varphi/2)$ and with the increase in the
energy the scattering becomes less symmetric till it becomes mainly forward at high energies. At higher energies
and larger $M,$ one obtains forward scattering with a relatively weak asymmetry 
$|\langle\varphi\rangle|\ll\,1 $ and small aperture $D_{\varphi}\ll\,1.$

\subsection{Diffraction gratings}

\begin{figure}[tbp]
\includegraphics*[width=0.4\textwidth]{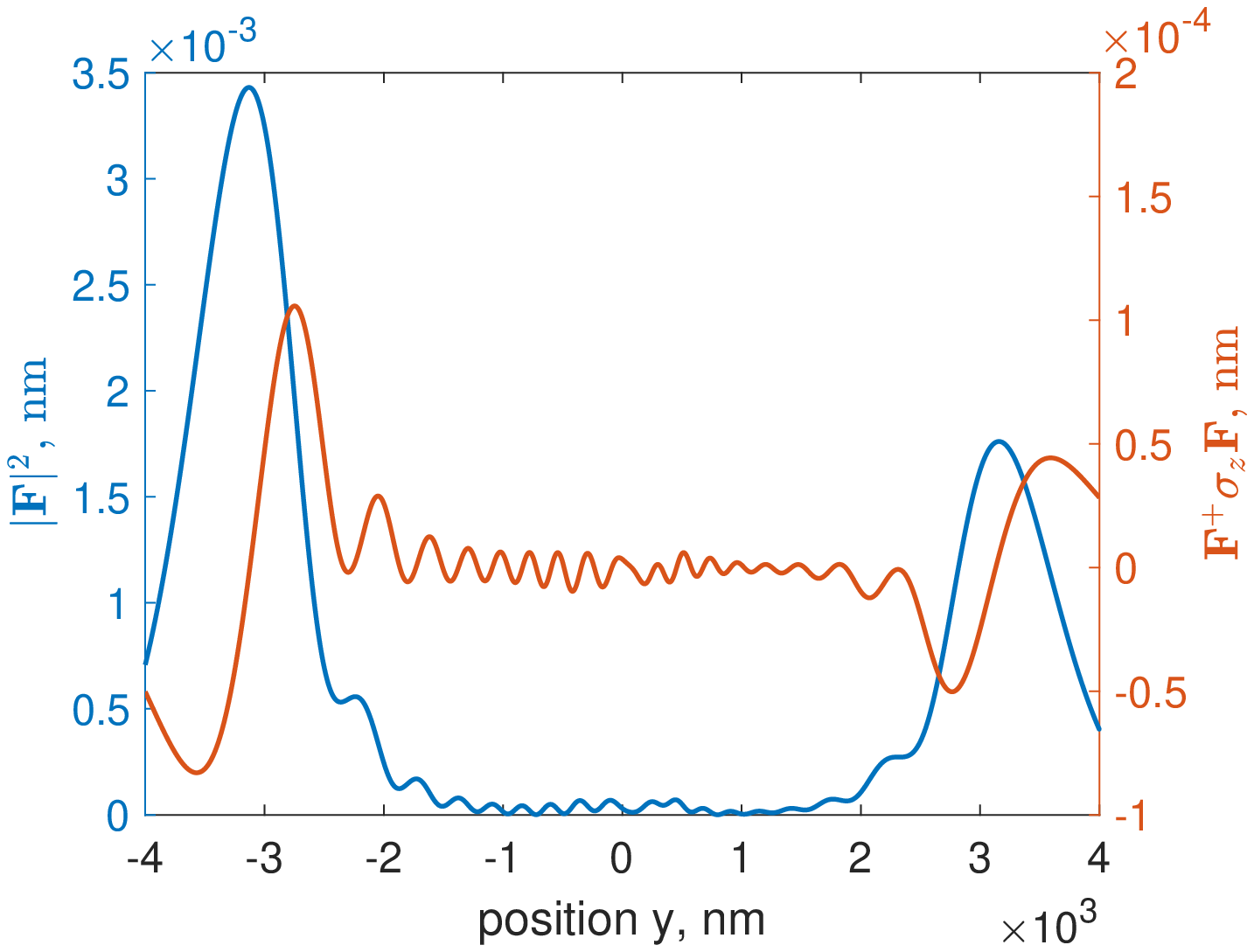}\\
\includegraphics*[width=0.4\textwidth]{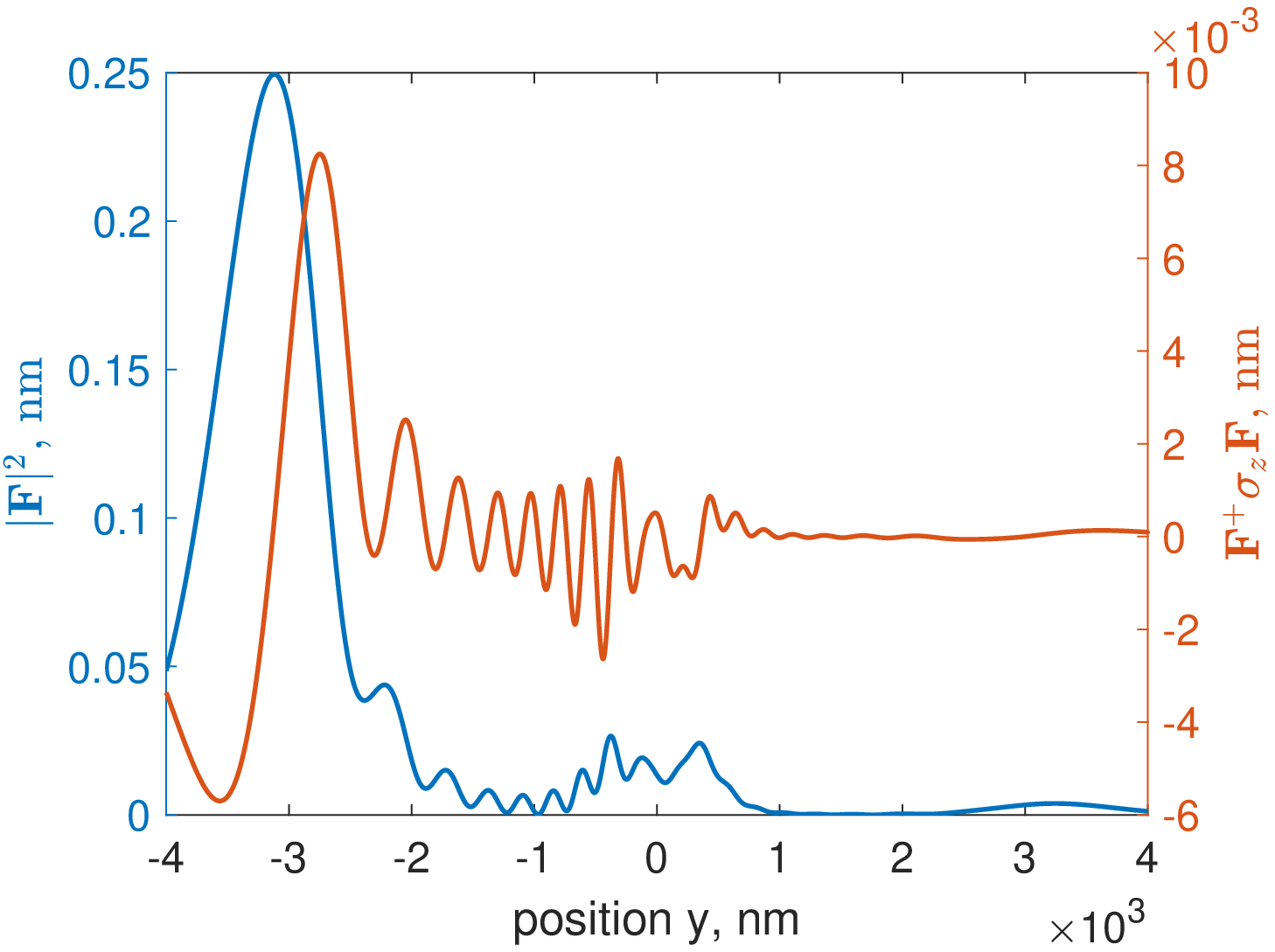}\\
\caption{Scattering probability and spin density produced by nanodots-formed diffraction grating with $N=10$ dots,
the interdot distance $d=150$ nm, and distance to the screen $L=2000$ nm. 
Upper panel corresponds to $R=5$ nm, lower panel corresponds to $R=15$ nm. 
Electron energy $\varepsilon=\Delta_{\rm m}/2=12.5$ meV, wavevector $k=5\times\,10^{-2}$ nm$^{-1},$ and the wavelength 
$\lambda=1.26\times\,10^{2}$ nm. Scattering angle $\varphi$ for the first principal maxima obtained with 
$d|\sin\varphi|=\lambda$ yields $|\sin\varphi|=\lambda/d=0.84$ and corresponds to position on 
the screen $|y|=L|\sin\varphi|/\sqrt{1-\sin^{2}\varphi}\approx\,3100$ nm, 
in agreement with the presented plots. Although Eqs. \eqref{eq:FF} and \eqref{eq:Fsigmaz} are not 
directly applicable here, they agree with these results demonstrating small $\sigma_{z}(y)$
near the principal peaks maxima.}
\label{fig:grating}
\end{figure}

Having discussed single-dot scattering, here we present in Fig. \ref{fig:grating} 
the numerical results for the density probability and
spin density produced by a diffraction grating. 
	
As shown in Fig. \ref{fig:grating}, the diffraction pattern consists of strong principal 
scattering peaks and of weak secondary intermediate
peaks, as predicted by the diffraction theory \cite{BornWolf}. As expected, the diffraction pattern is 
strongly asymmetric with $\langle\varphi\rangle<0,$ 
as can be understood from Figs. \ref{fig:3Dplots} and \ref{fig:1Dexamples}. In addition, we see that the $\sigma_{z}(y)$ 
spin density is small but not zero, as expected from the discussion above when for a single scatterer $\sigma_{z}(y)=0.$ 
Figure \ref{fig:grating} shows that with the given gratings geometry
one can achieve the spin polarization $\sigma_{z}(y)/\left|\mathbf{F}(y)\right|^{2}\le 0.1.$ 
This is a result of interference
of the waves scattered by different angles $\varphi_{i}$, similar to the effects observed in the 
scattering of bunches of ultrafast electrons in solids \cite{Michalik2008}.

\section{Conclusions}

We studied cross-section and diffraction patterns of electron scattering
by magnetic nanodots and their diffraction gratings on the surface of a topological insulator
with spin-momentum locking. {For a single nanomagnet, we considered analytically and numerically 
various scattering regimes in terms of the electron energy and
nanodot size and magnetization and demonstrated that they can be universally described by two dimensionless system
parameters.} The scattering probability is usually angle-asymmetric, presenting its qualitative 
feature due to the spin-momentum locking as can occur in
a broad interval of the scattering angles. It becomes angle-symmetric 
(i) at high energies, where it is concentrated in a narrow angle and (ii)
in the energy-independent Born approximation leading to the universal 
broad scattering probability distribution. {We demonstrated that the spins of scattered electrons remain 
parallel to the surface of the topological insulator.}  Next, we obtained the corresponding patterns of the scattering 
by diffraction gratings. In qualitative contrast to single scatterers, diffraction gratings 
produce nonzero {perpendicular to the surface} spin component of the scattered electrons. 

These results can be applied for the design of magnetization patterns such as
arrays of magnetic quantum dots or magnetization lattices {of nanomagnets of the size between 10 and 100 nm \cite{Cowburn1999}}
to produce in a controllable way spin and charge currents and densities at the surfaces of
topological insulators. This approach can be used for studies of spin torques 
\cite{Mellnik2014,Ndiaye2017,Han2017,Ghosh2017,Bondarenko2017,Mahendra2018,Moghaddam2020,Han2021}
produced on the magnetized quantum dots by
scattered electrons. Another application can be related to electron interferometry and 
holography of magnetic structures and nonuniform magnetic fields \cite{Fukuhara1983,Tonomura1987} 
providing detailed information about
magnetization patterns by visualization of the phase of the electron wavefunction. 

\section*{Acknowledgements}

This work was supported by the National Science Center in Poland as a
research project No. DEC-2017/27/B/ST3/02881. The work of E.S. 
is financially supported through Grants No. PGC2018-101355-B-I00 and 
No. PID2021-126273NB-I00 funded by MCIN/AEI/10.13039/501100011033 and 
by ERDF “A way of making Europe,” and by the Basque Government through 
Grants No. IT986-16 and No. IT1470-22.

\end{document}